\documentclass[letterpaper,preprintnumbers,prd,twocolumn,nofootinbib,nobibnotes,showpacs]{revtex4}
\usepackage{amsfonts}
\usepackage{mathrsfs}
\usepackage{epsfig}
\usepackage{graphicx}%
\usepackage{dcolumn}
\usepackage{amsmath}

\makeatletter
\def\btt#1{\texttt{\@backslashchar#1}}%
\DeclareRobustCommand\bblash{\btt{\@backslashchar}}%
\makeatother
\begin{document}

\title{Cosmological Models with Lagrange Multiplier Field}
\author{Changjun Gao$^{1,2}$}\email{gaocj@bao.ac.cn}\author{Yan Gong$^{1,3}$}\email{gongyan@bao.ac.cn}\author{Xin Wang$^{1,3}$}\email{wangx@bao.ac.cn}\author{Xuelei Chen$^{1,2,4}$}\email{xuelei@cosmology.bao.ac.cn}\affiliation{$^{1}$The
National Astronomical Observatories, Chinese Academy of Sciences, Beijing 100012, China}
\affiliation{{$^{2}$Kavli Institute for Theoretical
Physics China, CAS, Beijing 100190, China }}
\affiliation{$^{3}$Graduate School of Chinese Academy of Sciences, Beijing 100049}
\affiliation{$^{4}$Center for High Energy Physics, Peking University, Beijing 100871, China}

\date{\today}

\begin{abstract}
We first consider the Einstein-aether theory with
a \emph{gravitational coupling} and a Lagrange multiplier field, and then
consider the non-minimally coupled quintessence field theory with
Lagrange multiplier field. We study the influence of the
Lagrange multiplier field on these models.
We show that the energy density evolution of the Einstein-aether
field and the quintessence field are significantly modified.
The energy density of the Einstein-aether is nearly
a constant during the entire history of the Universe.
The energy density of the quintessence field can
also be kept nearly constant in the matter dominated Universe, or
even exhibit a phantom-like behavior for some models. This
suggests a possible dynamical origin of the
cosmological constant or dark energy. Further more, for the
canonical quintessence
in the absence of gravitational coupling, we find that the quintessence
scalar field can play the role of cold dark matter with the introduction of
a Lagrange multiplier field. We conclude that the Lagrange
multiplier field could play a very interesting and important role in
the construction of cosmological models.
\end{abstract}


\maketitle

\section{Introduction}
Ever since the discovery of cosmic acceleration
in 1998 \cite{perl:99, Riess:98}, the dark energy has
remained a fundamental mystery, both for
its unexpectedly small but non-zero value, and for
the apparent coincidence of its present density
being approximately that of the other components. Many attempts
have been made to address these problems, including, e.g.,
the cosmological constant, quintessence \cite{caldwell:1998},
phantom \cite{caldwell:1999}, quintom \cite{feng: 2004},
K-essence \cite{k-essence:2000},
holographic dark energy \cite{li:2004, gao:2009},
agegraphic dark energy \cite{cai:2007,wei:2008},
modified gravity \cite{many}, Einstein-aether theory \cite{ted:2001, li:2009},
and so on. It is however fair to say
that none of these interesting ideas has emerged to be the clear front
runner. Many of these are only toy models which needs to be further developed
to be decisively tested by observations \cite{cald:2009}.

In this paper, we start our investigation with the
Einstein-aether field theory\cite{ted:2001, li:2009}.
The Einstein-aether theory is an extension of the general relativity theory
which incorporates a dynamical unit timelike vector
field $A^{\mu}$. The presence of this field
breaks the local Lorentz symmetry down to
a three dimensional rotation subgroup. Direct coupling of the aether to
the matter would violate local Lorentz symmetry yet preserve
diffeomorphism invariance. This paper assumes that the aether
does not couple directly to the matter.

The Lagrangian density of the Einstein-aether theory is given by \cite{ted:2001}
\begin{eqnarray}
\label{eq:ted}
\mathscr{L}&=&\sqrt{-g}\left[K+\lambda\left(g_{\mu\nu}A^{\mu}A^{\nu}+m^2\right)\right]\;,
\end{eqnarray}
where
\begin{eqnarray}
{K}&=&{K}^{\alpha\beta}_{\ \ \ \gamma\sigma}\nabla_{\alpha}A^{\gamma}\nabla_{\beta}A^{\sigma}\;,\nonumber\\
{K}^{\alpha\beta}_{\ \ \
\gamma\sigma}&=&c_1g^{\alpha\beta}g_{\gamma\sigma}+c_2\delta_{\gamma}^{\alpha}\delta_{\sigma}^{\beta}
+c_3\delta_{\sigma}^{\alpha}\delta_{\gamma}^{\beta}\;.
\end{eqnarray}
Here $c_i$ and $m$ are constants, $g$ is the trace of the metric
tensor $g_{\mu\nu}$. We emphasize that $\lambda$ is not a constant,
but a Lagrange multiplier field or auxiliary field. The
Einstein-aether is similar to the vector tensor gravity theories
studied by Will and Nordvedt \cite{will:1972}. However, due to the
presence of the Lagrange multiplier field, there is a crucial
difference: the vector field is constrained to have unit norm.

Carroll et al. \cite{carroll:2009,carroll:20091} showed that the
models with the generic kinetic term given above are plagued by either ghosts
or tachyon, therefore physically unacceptable. They found that
only the timelike Sigma-model \cite{carroll:2009,carroll:20091}:
\begin{eqnarray}
\label{eq:sigma}
\mathscr{L}&=&\sqrt{-g}\left[-\nabla_{\mu}A_{\nu}\nabla^{\mu}A^{\nu}-R_{\mu\nu}A^{\mu}A^{\nu}
\right.\nonumber
\\
&&\left.+\lambda\left(g_{\mu\nu}A^{\mu}A^{\nu}+m^2\right)\right]\;,
\end{eqnarray}
is well-defined and stable. In this paper, we shall modify the fixed-norm constraint from the unit norm to
an environment-dependent norm. In detail, we modify the constraint
condition from
 \begin{eqnarray}
 \label{eq:unit}
g_{\mu\nu}A^{\mu}A^{\nu}+m^2=0\;,
\end{eqnarray}
to
 \begin{eqnarray}
G_{\mu\nu}A^{\mu}A^{\nu}+m^2=0\;,
\end{eqnarray}
where $G_{\mu\nu}=R_{\mu\nu}-\frac{1}{2}Rg_{\mu\nu}$
is the Einstein tensor. We will shortly find that this modification is not trivial. In fact,
in our model the energy density of the aether is nearly a constant during the
entire history of the Universe, while the energy density of the
unit norm aether is proportional to $H^2$ ($H$ is the Hubble
parameter) \cite{carroll:2004}.

Inspired by this interesting result, we also consider the Lagrange
multiplier field in the non-minimally coupled quintessence theory.
After we completed the discussions on the non-minimally coupled
quintessence case, we learned that the Lagrange multiplier field
had been introduced earlier by Mukhanov and Brandenberger in
\cite{muk:1992}. We find that the energy density of the
quintessence can also be kept at nearly constant value in the
matter dominated Universe. This suggests a way of finding a
dynamical origin of the cosmological constant. Furthermore, by
exploring a minimally coupled quintessence field with the Lagrange
multiplier field, we find that the quintessence behaves as
pressureless matter, which enables it to be also a promising
candidate of the cold dark matter. It is also found that the
adiabatic sound speed and the rest-frame sound speed of the
quintessence is exactly zero. In this respect, our conclusion is
consistent with the investigation of Ref.~\cite{lim:2010}.
Different from Ref.~\cite{lim:2010}, we also investigated the
Lagrange multiplier field in the Einstein-aether cosmology with
the modified constraint conditions.

We note that many new results have appeared ever since the first
version of this work is present in the preprint form
\cite{manyy1,manyy2,manyy3,manyy4,manyy5,manyy6,manyy7,manyy8,manyy9}.
Capozziello et al. \cite{manyy1} studied the scalar-tensor theory,
k-essence and modified gravity with Lagrange multiplier
constraint. They conclude that the well-known mathematical
equivalence between scalar theory and $f(R)$ gravity is broken due
to the presence of constraint. Then the important conclusion leads
us to look for the viable gravity theory among vast originally
non-realistic ones. Using the scalar and vector Lagrange
multiplier method, Nojiri et al. \cite{manyy2} proposed a class of
covariant gravity theories which have nice ultraviolet behavior
and are potentially renormalizable. Cai and Saridakis
\cite{manyy3} investigated the cyclic and singularity-free
cosmological solutions using the Lagrange multiplier method. They
showed that the realization of cyclicity and the avoidance of
singularities is very straightforward in this scenario. Feng and
Li \cite{manyy4} calculated the primordial curvature perturbation
for the curvaton model in the presence of a Lagrange multiplier
field. On the other hand, Kluson \cite{manyy5} developed the
Hamiltonian formalism for the Lagrange multiplier modified
gravity. For more great detail, we prefer the reader to the nice
review paper Ref.~\cite{manyy7}.

The paper is organized as follows. In section II, we will
investigate the cosmic evolution and observational constraints
on the Einstein-aether with the modified norm-fixing condition.
In section III, we investigate the cosmic evolution of the
non-minimally coupled quintessence in the presence of a
Lagrange multiplier field. In section VI, we show the minimally
coupled quintessence behaves as pressureless matter in the presence
of a Lagrange multiplier field, and calculate both the adiabatic
sound speed and the rest frame sound speed. Section VI discusses the results
and concludes. We shall use the system of units with $G=c=\hbar=k=1$
and the metric signature $(-,\ +,\ +,\ +)$ throughout the paper.

\section{Cosmological constant from Einstein-aether}
\subsection{equation of motion}
We start with the following form of action:
\begin{eqnarray}
\label{eq:our}
\mathscr{L}&=&\sqrt{-g}\left[\left(\nabla_{\mu}A^{\mu}\right)^2+\lambda\left(G_{\mu\nu}A^{\mu}A^{\nu}+m^2\right)\right]\;.
\end{eqnarray}
The equation of motion obtained by varying the action with respect to $\lambda$ enforces the norm-fixing  constraint
 \begin{eqnarray}
 \label{eq:cons}
G_{\mu\nu}A^{\mu}A^{\nu}+m^2=0\;.
\end{eqnarray}
The equation of motion is obtained by varying the action
with respect to $A^{\mu}$,
\begin{eqnarray}
\label{eq:eom}
\nabla_{\mu}\left(\nabla_{\alpha}A^{\alpha}\right)=\lambda G_{\mu\alpha}A^{\alpha}\;.
\end{eqnarray}
Multiplying Eq.~(\ref{eq:eom}) with $A^{\mu}$ on both sides, and use the
constraint Eq.~(\ref{eq:cons}), we obtain
\begin{eqnarray}
\label{eq:L}
\lambda=-\frac{1}{m^2}A^{\mu}\nabla_{\mu}\left(\nabla_{\alpha}A^{\alpha}\right)\;.
\end{eqnarray}
The energy momentum tensor is obtained by the action varying with
respect to $g^{\mu\nu}$ \cite{pic:2009}:

\begin{eqnarray}
\label{eq:stress}
T_{\mu\nu}^{A}&=&-2\left[{\textstyle\frac12}g_{\mu\nu}\left(\nabla_{\alpha}A^{\alpha}\right)^2+g_{\mu\nu}A^{\alpha}\nabla_{\alpha}\left(\nabla_{\beta}A^{\beta}\right)
\right.\nonumber
\\
&&\left.-2A_{(\mu}\nabla_{\nu)}\left(\nabla_{\alpha}A^{\alpha}\right)\right]
+\lambda\left[2\nabla_{\alpha}\nabla_{(\mu}\left(A_{\nu)}A^{\alpha}\right)\right.\nonumber
\\
&&\left.-\nabla_{\alpha}\nabla_{\beta}\left(A^{\alpha}A^{\beta}\right) g_{\mu\nu}-\nabla^2 \left(A_{\mu}A_{\nu}\right)\right.\nonumber
\\
&&\left.-4A^{\alpha}R_{\alpha ( \mu} A_{\nu )}+R_{\mu\nu}A_{\alpha}A^{\alpha}+RA_{\mu}A_{\nu}\right.\nonumber
\\
&&\left.-\nabla_{\mu}\nabla_{\nu}\left(A^{\alpha}A_{\alpha}\right)+g_{\mu\nu}\nabla^2\left(A_{\alpha}A^{\alpha}\right)\right.\nonumber
\\
&&\left.+g_{\mu\nu}\left(G_{\mu\nu}A^{\mu}A^{\nu}+m^2\right)\right]\;.
\end{eqnarray}
We note that $G_{\mu\nu}$ vanishes in the Minkowski spacetime. It is then follows that $m=0$ from Eq.~(\ref{eq:cons}) and the model is reduced to
\begin{eqnarray}
\label{eq:mink}
\mathscr{L}&=&\sqrt{-g}\left(\nabla_{\mu}A^{\mu}\right)^2\;,
\end{eqnarray}
which has been studied in Ref.~\cite{pic:2009}.

\subsection{cosmological solution}
Observations reveal that the spatial geometry of the
Universe is almost flat, so below we consider the spatially flat
Friedmann-Robertson-Walker (FRW) Universe:
\begin{eqnarray}
ds^2&=&-dt+a\left(t\right)^2\left(dr^2+r^2d\Omega^2\right)\;,
\end{eqnarray}
with $a(t)$ the scale factor. For such a metric, the vector must respect
spatial isotropy, at least at the background level. Thus the only non-vanishing
component of the vector should be the timelike component. Using the norm-fixing
constraint Eq.~(\ref{eq:cons}), the components of the vector field are simply
\begin{eqnarray}
A^{\mu}&=&\left(\frac{m}{\sqrt{3}H},\ \ 0,\ \ 0,\ \ 0\right)\;,
\end{eqnarray}
where $H$ is the Hubble parameter. From Eq.~(\ref{eq:L}) we then find
the Lagrange multiplier field $\lambda$ must satisfy
 \begin{eqnarray}
\lambda&=&\frac{1}{3H}\frac{d}{dt}\left(\frac{\dot{H}}{H^2}\right)\;.
\end{eqnarray}
where the dot above represents the derivative with respect to
cosmic time $t$.  Using the above two equations, we find from
Eq.~(\ref{eq:stress}) that the energy density and pressure of the
vector field (and the Lagrange multiplier field) are given by
 \begin{eqnarray}
\rho_A&=&\frac{m^2}{3}\left(3-\frac{\dot{H}}{H^2}\right)^2\;,
\end{eqnarray}
and
\begin{eqnarray}
p_A&=&-\rho_A+\frac{2m^2}{9}\left(3-\frac{\dot{H}}{H^2}\right)\left(\frac{\dot{H}}{H^2}\right)^{\cdot}\;,
\end{eqnarray}
respectively. It can be easily shown that the energy density and
pressure satisfy the conservation equation
\begin{eqnarray}
\dot{\rho}_A+3H\left(\rho_A+p_A\right)=0\;,
\end{eqnarray}
which is just an expression of the conservation law for energy
momentum tensor

\begin{eqnarray}
T_{\mu\nu}^{A;\mu}=0\;.
\end{eqnarray}

On the other hand, the Einstein equations tell us that
 \begin{eqnarray}
 \label{eq:H}
\frac{\dot{H}}{H^2}&=&-\frac{3}{2}\left(1+w\right)\;.
\end{eqnarray}
Here $w\equiv{p}/{\rho}$ and $\rho,\ p$ represent the total energy
density and total pressure of the comic fluids (\ including aether
field\ ). Then the energy density can also be written as
 \begin{eqnarray}
\rho_A=\frac{m^2}{3}\left[3+\frac{3}{2}\left(1+w\right)\right]^2\;.
\end{eqnarray}
For the radiation dominated Universe, we have $w\simeq 1/3$. For
the matter dominated Universe, we have $w\simeq 0$. Hence we have
$\rho_A=8.33\  m^2$ and $\rho_A=6.75\ m^2$ respectively. In both
cases, the energy density remains nearly constant. For the
present-day Universe, we have $w\simeq {-0.75}$ such that
$\rho_A\simeq3.79\ m^2.$ In all of these cases, \emph{the energy
density of the
 aether field behaves as a constant during the whole history of the Universe.
 } It follows immediately that the equation of state of the aether
is $w_A\simeq -1$. To show the point in great detail, let's
numerically investigate the Friedmann equation in the following.

Taking into account radiation and matters (including ordinary
matter and dark matter), we obtain the Friedmann equation

 \begin{eqnarray}
 \label{eq:Fr}
3H^2=8\pi\left[\frac{\rho_{r0}}{a^4}+\frac{\rho_{m0}}{a^3}+\frac{m^2}{3}\left(3-\frac{\dot{H}}{H^2}\right)^2\right]\;.
\end{eqnarray}
Here $\rho_{r0}$ and $\rho_{m0}$ represent the energy densities of
radiation and matter, respectively, in the present-day universe.

Define
\begin{eqnarray}
x\equiv\ln a\;.
\end{eqnarray}
The Friedmann equation can be rewritten as
 \begin{eqnarray}
 \label{eq:Fr2}
3H^2=8\pi\left[{\rho_{r0}}e^{-4x}+{\rho_{m0}}e^{-3x}+\frac{m^2}{3}\left(3-\frac{1}{H}\frac{d{H}}{dx}\right)^2\right]\;.
\end{eqnarray}
Let
\begin{eqnarray}
h\equiv\frac{H}{H_0}\;,\ \
\Omega_{r0}\equiv\frac{\rho_{r0}}{\rho_0}\;,\ \
\Omega_{m0}\equiv\frac{\rho_{m0}}{\rho_0}\;,\ \
\zeta^2\equiv\frac{8\pi m^2}{9H_0^2}\;,\ \  \;,
\end{eqnarray}
then we have
 \begin{eqnarray}
 \label{eq:Fr2}
h^2={\Omega_{r0}}e^{-4x}+{\Omega_{m0}}e^{-3x}+\zeta^2\left(3-\frac{1}{h}\frac{d{h}}{dx}\right)^2\;.
\end{eqnarray}
Here $H_0, \ \rho_0$ are the Hubble parameter, total energy
density for the present-day Universe. $\zeta$ is a dimensionless
free parameter. The standard cosmological model, e.g. as in
Komatsu et al. \cite{Komatsu5yrWMAP}, predicts the present matter
density ratio $\Omega_{m0}=0.25$ and radiation density ratio
$\Omega_{r0}=8.1\cdot 10^{-5}$. Using this result and taking
$\zeta=0.25$, we plot the dimensionless Hubble parameter $h$ via
redshifts $z$ in Fig.~1 for our model and the standard $\Lambda
\textrm{CDM}$ model. We conclude from the figure that it is
consistent with the observations.

In order to show the aether model can lead to the cosmic
acceleration, we plot the evolution of deceleration parameter $q$:
 \begin{eqnarray}
q\equiv-\left(1+\frac{d\ln h}{dx}\right)\;,
\end{eqnarray}
in Fig.~2. We find the model predicts nearly the same transition
redshift $z_T=0.8$ for the Universe from deceleration to
acceleration as the standard $\Lambda \textrm{CDM}$ model.

\begin{figure}
\includegraphics[width=6.5cm]{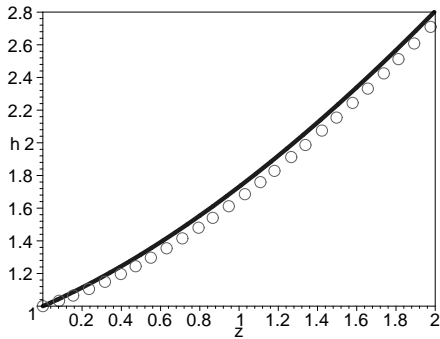}
\caption{The dimensionless Hubble parameter $h$ with redshift $z$.
The circled line is for the standard $\Lambda \textrm{CDM}$ model.
The solid line is for the aether model.} \label{fig:hz}
\end{figure}

\begin{figure}
\includegraphics[width=6.5cm]{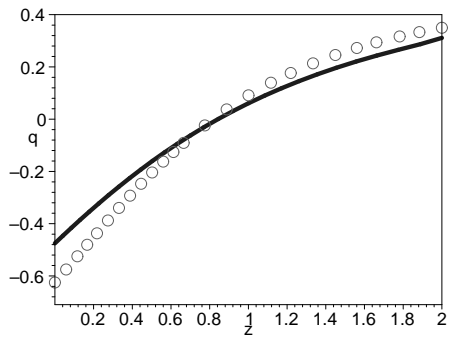}
\caption{The deceleration parameter $q$ with redshift $z$. The
circled line is for the standard $\Lambda \textrm{CDM}$ model. The
solid line is for the aether model.} \label{fig:qz}
\end{figure}

\begin{figure}
\includegraphics[width=6.5cm]{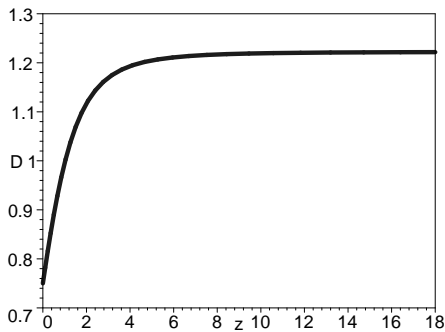}
\caption{The dimensionless energy density $D$ with redshift $z$.
At redshifts greater than $4$, the energy density of aether field
is nearly a constant.} \label{fig:dz}
\end{figure}

In Fig.~3, we plot the dimensionless energy density (from
Eq.~(\ref{eq:Fr2}))

\begin{eqnarray}
 D=\zeta^2\left(3-\frac{1}{h}\frac{d{h}}{dx}\right)^2\;,
\end{eqnarray}
for the aether field with the redshift. It shows the energy
density of aether field is nearly a constant at redshifts greater
than $4$.

Carroll et al \cite{carroll:2004} showed that the energy density
of aether field is proportional to $H^2$ in the absence of
gravitational coupling. The reason for this difference could be
understood as follows: for the usual Einstein-aether theory, the
norm constraint condition Eq.~(\ref{eq:unit}) gives (by order) in
flat FRW Universe \cite{carroll:2004}
\begin{eqnarray}
A^{\mu}\sim m\;.
\end{eqnarray}
The norm of the vector is fixed as a constant. On the other hand,
dimensional analysis suggests that we must have
\begin{eqnarray}
\nabla_{\mu}\sim H\;.
\end{eqnarray}
Therefore, we conclude that the energy density is
\begin{eqnarray}
\nabla_{\mu}A^{\mu}\nabla_{\nu}A^{\nu}\sim H^2\;,
\end{eqnarray}
which is consistent with the calculation of
Ref.~\cite{carroll:2004}. For our Einstein-aether field, from the
constraint equation Eq.~(\ref{eq:cons})
\begin{eqnarray}
A^{\mu}\sim \frac{m}{H}\sim\frac{m}{\sqrt{\rho}}\;,
\end{eqnarray}
where $\rho$ is the total energy of the Universe.
It is apparent that the norm of our vector is environment dependant.
The energy density turns out to be
\begin{eqnarray}
\nabla_{\mu}A^{\mu}\nabla_{\nu}A^{\nu}\sim H^2\cdot H^{-2}\simeq {\rm const.}\;
\end{eqnarray}
As another example, we have checked with detailed calculation that for
the Einstein-aether theory with the Lagrangian density
\begin{equation}
\label{eq:our2}
\mathscr{L}=\sqrt{-g}\left[G_{\mu\nu}\nabla^{\mu}A^{\alpha}\nabla^{\nu}A_{\alpha}
+\lambda\left(G_{\mu\alpha}G^{\alpha}_{\nu}A^{\mu}A^{\nu}+m^2\right)\right]\;,
\end{equation}
the energy density of aether in this model is also kept at nearly
constant level. However, the corresponding Einstein equations
involves derivatives higher than second order.

\section{Cosmological constant and Phantom from quintessence}

\subsection{equation of motion}
Inspired by the dynamical behavior of the Einstein-aether theory
with the Lagrange multiplier field, we consider its role
in quintessence field theory.
For our purpose, here we consider the following model:
\begin{equation}
\label{eq:qui}
\mathscr{L}=-\sqrt{-g}\left[\frac{1}{2}G_{\mu\nu}\nabla^{\mu}\phi\nabla^{\nu}\phi+\lambda\left(\frac{1}{2}R\phi+\frac{m^2}{2}\right)\right]\;
\end{equation}
where $m$ is a constant, $R$ is the Ricci scalar. The case where $\lambda$ is
a constant has been investigated extensively. However,
we emphasize that $\lambda$ is \emph{not} a constant,
but a Lagrange multiplier or auxiliary \emph{field}. This brings interesting new
possibilities.

The equation of motion for $\lambda$ and $\phi$ are given by
\begin{eqnarray}
\label{eq:ql}
R\phi+m^2=0\;,\\
\label{eq:q2}
G_{\mu\nu}\nabla^{\mu}\nabla^{\nu}\phi-\frac{\lambda}{2}R=0\;,
\end{eqnarray}
and the energy-momentum tensor takes the
form \cite{sushkov:2009,ford:1987,uzan:1999}
\begin{eqnarray}
T_{\mu\nu}&=&-{\textstyle\frac12}\nabla_\mu\phi\,\nabla_\nu\phi\,R
+2\nabla_\alpha\phi\,\nabla_{(\mu}\phi R^\alpha_{\nu)}
\nonumber\\
&&
+\nabla^\alpha\phi\,\nabla^\beta\phi\,R_{\mu\alpha\nu\beta}+\nabla_\mu\nabla^\alpha\phi\,\nabla_\nu\nabla_\alpha\phi
\nonumber\\
&&
-\nabla_\mu\nabla_\nu\phi\,\square\phi-{\textstyle\frac12}(\nabla\phi)^2
G_{\mu\nu}
\nonumber\\
&&
+g_{\mu\nu}\big[-{\textstyle\frac12}\nabla^\alpha\nabla^\beta\phi\,\nabla_\alpha\nabla_\beta\phi
+{\textstyle\frac12}(\square\phi)^2
\nonumber\\
&& -\nabla_\alpha\phi\,\nabla_\beta\phi\,R^{\alpha\beta}
\big]-{\textstyle\frac12}\lambda m^2g_{\mu\nu}+\lambda G_{\mu\nu}\phi\nonumber\\&&+\lambda g_{\mu\nu}\nabla^2\phi-\lambda \nabla_{(\mu}\nabla_{\nu)}\phi\;.
\end{eqnarray}

\subsection{cosmological solution}
With the flat FRW background model, we obtain
\begin{eqnarray}
\label{eq:q11}
\phi&=&-\frac{m^2}{{R}}=\frac{m^2}{{{6\dot{H}+12H^2}}}\;,\\
\label{eq:q22}
\lambda&=&\frac{2}{m^2}\phi\left[3H^2\ddot{\phi}+3H\left(2\dot{H}+3H^2\right)\dot{\phi}\right]\;.
\end{eqnarray}

From the energy-momentum tensor, the energy density of scalar is
 \begin{equation}
 \rho_{\phi}=\frac{9}{2}H^2\dot{\phi}^2+3\lambda H^2\phi+3\lambda H\dot{\phi}+\frac{1}{2}\lambda m^2
 \end{equation}
Using Eq.~(\ref{eq:q11}) and Eq.~(\ref{eq:q22}), and assuming the total equation of state $w$ is a constant, we find the energy density is
\begin{eqnarray}
\label{eq:qd}
\rho_{\phi}=18{m^4}\cdot\frac{\left(1+w\right)\left(1-w-w^2\right)}{\left(1-3w\right)^3}\;.
\end{eqnarray}
We have $w\simeq0$ for the matter dominated Universe.
The corresponding energy density is
\begin{eqnarray}
\rho_{\phi}=18m^4\;.
\end{eqnarray}
This density does not vary as the Universe expands,
so \emph{the field behaves as a cosmological constant}.

For the radiation dominated case, we have $n\simeq1/2$. It is apparent that
Eq.~(\ref{eq:qd}) is divergent. This is not surprising, since
$R=0$ in radiation dominated Universe.
We should look for the energy density
from the Lagrangian. Ref.~{\cite{gao:2010}}
showed that the energy density scales as
\begin{eqnarray}
\rho_{\phi}=\frac{\rho_0}{a^2}\;,
\end{eqnarray}
in the radiation dominated epoch, with $\rho_0$ an integration constant.
Then the field has the equation of state $w=-1/3$.
The energy density evolution given in Eq.~(\ref{eq:qd}) is
different from the usual non-minimally coupled quintessence field. For the present-day Universe, we have
$w\simeq {-0.75}$ such that $\rho_{\phi}\simeq0.16\ m^2.$

The reason for this difference could be understood as follows.
The fixed-norm constraint Eq.~(\ref{eq:ql}) (also Eq.~(\ref{eq:q11})) tells us
\begin{eqnarray}
\phi\sim\frac{m^2}{H^2}\sim\frac{m^2}{{\rho}}\;.
\end{eqnarray}
In other words, the strength of the field is determined by
the energy density of the background matter source.
In the usual quintessence theory, the strength of the field
can be arbitrarily large. However,
we have $\nabla_{\mu}\sim H$ and $G_{\mu\nu}\sim H^2$, so
\begin{eqnarray}
G_{\mu\nu}\nabla^{\mu}\phi\nabla^{\mu}\phi\sim H^2\cdot H^2\cdot H^{-4} \sim
{\rm const.}\;
\end{eqnarray}

Another interesting model is
\begin{eqnarray}
\mathscr{L}&=&-\sqrt{-g}\left[\frac{1}{2}\nabla_{\mu}\phi\nabla^{\mu}\phi+\lambda\left(\frac{1}{2}R\phi+\frac{m^2}{2}\right)\right]\;,
\end{eqnarray}
in this case the scalar field behaves as ``phantom'',
with equation of state $w<-1$. This
can be seen by the following reasoning: the norm-fixing constraint yields
\begin{eqnarray}
\phi\sim\frac{m}{H^2}\;.
\end{eqnarray}
So the energy density is
\begin{eqnarray}
\nabla_{\mu}\phi\nabla^{\mu}\phi\sim H^2\cdot{H^{-4}}\sim H^{-2}\;.
\end{eqnarray}
The energy density increases with the expansion of the Universe, i.e.
it behaves as a phantom. By using the Lagrange Multiplier field, one may
construct the dark energy model which exhibit the desired behaviors.

\section{cold dark matter from quintessence}
\subsection{the theory}
In Section II and Section III, we have shown that the energy density
of the Einstein-aether and the quintessence can be kept at nearly a
constant value in the evolution of the Universe by employing a
Lagrange multiplier field, which points to a dynamical origin of
cosmological constant. In this section, we shall show that with
proper choice of the Lagrange multiplier field, the cold dark matter
behavior can also be obtained.

Note that the usual quintessence field can have zero pressure for particular
potential forms. However, the sound speed for the canonical quintessence field
is 1, so it is not a true cold dark matter. Here we show how with the help
of Lagrange multiplier field, we can construct true cold dark matter with the
quintessence field.

We consider the following Lagrangian density:
\begin{eqnarray}
\mathscr{L}&=&-\lambda\sqrt{-g}\left[\frac{1}{2}\nabla_{\mu}\phi\nabla^{\mu}\phi+V\left(\phi\right)\right]\;,
\end{eqnarray}
where $\lambda$ is a Lagrange multiplier field,
and $V\left(\phi\right)$ is the scalar potential.
The equation of motion for $\lambda$ and $\phi$ are given by
\begin{eqnarray}
\label{eq:dml}
\frac{1}{2}\nabla_{\mu}\phi\nabla^{\mu}\phi+V\left(\phi\right)=0\;,\\
\label{eq:dmphi}
\nabla_{\mu}\left(\lambda\nabla^{\mu}\phi\right)-\lambda V^{'}=0\;,
\end{eqnarray}
Here prime denotes the derivative with respect to $\phi$.
The energy-momentum tensor is
\begin{eqnarray}
\label{eq:3stress}
T_{\mu\nu}=\lambda\left[\nabla_{\mu}\phi\nabla_{\nu}\phi-\frac{1}{2}g_{\mu\nu}\nabla_{\alpha}\phi\nabla^{\alpha}\phi-g_{\mu\nu}V\right]\;.
\end{eqnarray}
For a flat FRW model, Eq.~(\ref{eq:dml}) and Eq.~(\ref{eq:dmphi}) reduce to
\begin{eqnarray}
\label{eq:dm11} \frac{1}{2}\dot{\phi}^2-V=0\;,
\end{eqnarray}
and
\begin{eqnarray}
\label{eq:dmphi2}
\ddot{\phi}+3H\dot{\phi}+V^{'}+{\lambda}^{-1}\dot{\lambda}\dot{\phi}=0\;.
\end{eqnarray}
Differentiating Eq.~(\ref{eq:dm11}) with respect to time $t$, we
obtain
  \begin{eqnarray}
V^{'}=\ddot{\phi}\;.
\end{eqnarray}
Substituting it into Eq.~(\ref{eq:dmphi2}), we obtain
 \begin{eqnarray}
\frac{d\dot{\phi}}{\dot{\phi}}+\frac{3}{2}\frac{da}{a}+\frac{1}{2}\frac{d\lambda}{\lambda}=0\;.
\end{eqnarray}
Solving the equation, we have
 \begin{eqnarray}
 \label{eq:lambda2}
\lambda=\frac{\rho_0}{a^3\dot{\phi}^2}\;,
\end{eqnarray}
with $\rho_0$ an integration constant. Using
Eq.~(\ref{eq:3stress}) and Eq.~(\ref{eq:lambda2}), we find that
the energy density evolves as
 \begin{eqnarray}
 \rho_{\phi}=\frac{\rho_0}{a^3}\;,
\end{eqnarray}
which is exactly the behavior of the cold dark matter.
We note that this expression of energy density is \emph{independent} of the
explicit form of the scalar potential.

\subsection{speed of sound}
A dark matter model will fit the CMB data only if the rest-frame
sound speed is indeed zero \cite{mydeg}. As this is such an
important condition, it is worthwhile to spend some time to derive
the sound speed from the perturbation equations directly. For this
purpose, we work in the Newtonian gauge. In the absence of
anisotropic stress, and for scalar perturbations, the perturbed
flat FRW metric can be written in the form
\begin{eqnarray}
ds^2=-\left(1+2\Phi\right)dt^2+a\left(t\right)^2
\left(1-2\Phi\right)dx_idx^i\;,
\end{eqnarray}
where $\Phi$ is the Newtonian potential.
For the scalar field and the
Lagrange multiplier field, we define the perturbation as
\begin{eqnarray}
\label{eq:p1}
\phi\left(t,
\vec{x}\right)&=&\phi_0\left(t\right)+\delta\phi\left(t,\vec{x}\right)\;,\\
\label{eq:p2}
\lambda\left(t,
\vec{x}\right)&=&\lambda_0\left(t\right)+\delta\lambda\left(t,\vec{x}\right)\;.
\end{eqnarray}
The perturbed energy-momentum tensor is
\begin{eqnarray}
\label{eq:pert}
\delta T_0^0&=&\delta \rho_{\phi}=\lambda\left(\dot{\phi_0}\dot{\delta
  \phi}-\Phi\dot{\phi_0}^2+V^{'}\delta\phi\right)+\rho_{\phi0}\delta\lambda/\lambda\;,\nonumber\\
\delta T_i^j&=&-\delta
p_{\phi}\left(t,\vec{x}\right)\delta_{i}^{j}=-\lambda\left(\dot{\phi_0}\delta\dot{\phi}-\Phi\dot{\phi_0}^2-V^{'}\delta\phi\right)
\delta_{i}^{j}\;,\nonumber\\
ik\delta T_0^i&=&ik\left(\rho_{\phi0}+p_{\phi0}\right)\delta
u^i_{\phi}\left(t,\vec{x}\right)={\rho_{\phi0}}\mathscr{V}\;.
\end{eqnarray}
It is well-known that both the adiabatic sound speed and the rest
frame sound speed (the sound speed for the fluid in its rest frame)
play important roles in the discussion of structure formation
theory. Here we work out both of these quantities explicitly. The
adiabatic sound speed squared is defined as \cite{bean:2004,kunz:2006}
\begin{eqnarray}
c_a^2\equiv\frac{\dot{p}_{\phi 0}}{\dot{\rho}_{\phi0}}.
\end{eqnarray}
Since the pressure of this quintessence is vanishing, we conclude that
\begin{eqnarray}
c_a^2=0.
\end{eqnarray}

The rest frame sound speed squared $\hat{c}_{s}^2$ of the scalar is
related to the pressure perturbation in the Newtonian gauge through
\cite{bean:2004,kunz:2006}
\begin{eqnarray}
\label{e:dp}\delta
p_{\phi}=\hat{c}_{s}^2\delta\rho_{\phi}+\frac{3aH}{k^2}
\left(\hat{c}_{s}^2-c_a^2\right){\rho_{\phi0}}\mathscr{V}\;. \label{eq:dpcs}
\end{eqnarray}
Taking into account of $c_a^2=0$, we can rewrite Eq.~(\ref{e:dp}) as
\begin{eqnarray}
\label{e:dp1}\delta
p_{\phi}=\hat{c}_{s}^2\left(\delta\rho_{\phi}+\frac{3aH}{k^2}{\rho_{\phi0}}\mathscr{V}\right)\;.
\end{eqnarray}
In order to calculate $\delta p_{\phi}$, we resort to the equation of
motion for the scalar field perturbations. To this end, we insert
the expansions Eq.~(\ref{eq:p1}-\ref{eq:p2}) into the Lagrangian
density, and expand to quadratic order
\begin{eqnarray}
\mathscr{L}\rightarrow\mathscr{L}_0+\mathscr{L}_1+\mathscr{L}_2\;,
\end{eqnarray}
where $\mathscr{L}_1$ and $\mathscr{L}_2$ are the Lagrangian terms
which are of linear and quadratic order in $\delta\phi$,
respectively. Variation of $\mathscr{L}_0$ give us the equation
of motions for the unperturbed $\phi$ and $\lambda$.
Variation of $\mathscr{L}_2$ give us the equation of motion
for $\delta\phi$ and $\delta\lambda$, respectively.
Variation of $\mathscr{L}_1$ does not yield new information.

Straightforward calculation  yields
\begin{eqnarray}
\mathscr{L}_2&=&-a^3\delta\lambda\left(\Phi\dot{\phi}_0^2-\dot{\phi}_0\delta\dot{\phi}+V^{'}\delta\phi\right)\nonumber\\&&
-2a^3\Phi\delta\lambda\left(-\frac{1}{2}\dot{\phi}^2+V\right)\nonumber\\&&
+2a^3\Phi\lambda\left(\Phi\dot{\phi}_0^2-\dot{\phi}_0\delta\dot{\phi}+V^{'}\delta\phi\right)\nonumber\\&&
-\lambda a^3\left[2\Phi\dot{\phi}_0\delta\dot{\phi}-\frac{1}{2}\delta\dot{\phi}^2+\frac{1}{2a^2}\left(\nabla\delta{\phi}\right)^2
\right.\nonumber
\\
&&\left.+\frac{1}{2}V^{''}\delta{\phi}^2\right]\;.
\end{eqnarray}
Variation of $\mathscr{L}_2$ with respect
to $\delta\lambda$ and taking into account of Eq.~(\ref{eq:dm11}), we obtain
\begin{eqnarray}
\Phi\dot{\phi}_0^2-\dot{\phi}_0\delta\dot{\phi}+V^{'}\delta\phi=0\;.
\end{eqnarray}
Comparing this with the second equation in Eqs.~(\ref{eq:pert}), it follows that
\begin{eqnarray}
\delta T_i^j&=&-\delta
p_{\phi}\left(t,\vec{x}\right)\delta_{i}^{j}=0\;.
\end{eqnarray}
We then conclude immediately that
\begin{eqnarray}
\hat{c}_{s}^2=0\;.
\end{eqnarray}
Therefore, for this model \emph{the pressure, the adiabatic sound
speed, and the rest frame sound speed all vanishes, which ensures
that the scalar field has every desirable property of the cold dark
matter.}

\section{conclusion and discussion}
In conclusion, we have introduced \emph{gravitational coupling} to the
usual Einstein-aether theory and \emph{Lagrange multiplier field} to
the scalar field quintessence theory. We find that by changing the
norm-fixing condition from unit norm in the usual Einstein aether theory
to an environment-dependent one, one can make the aether density stay constant
as the Universe expands.

As for the case of scalar field quintessence, by introducing the
Lagrange multiplier field, the strength of the quintessence
field can be related to the background matter source, and the density can also
be kept as nearly constant during the cosmic expansion. This suggests a possible
dynamical origin of cosmological constant. Alternatively, we also find an
example in which the quintessence with Lagrange multiplier field
behaves as phantom.

Furthermore, with the help of a Lagrange multiplier field,
we also proposed a way to generate cold dark matter from quintessence.
It is found that the pressure, the adiabatic sound speed and the
rest-frame sound speed of the quintessence is exactly zero.
These properties enable the scalar field to be a potential
candidate for cold dark matter.

\acknowledgments

We sincerely thank the anonymous reviewer for the expert and
insightful comments, which have certainly improved the paper
significantly. This work is supported by the National Science
Foundation of China under the Key Project Grant 10533010, Grant
10575004, Grant 10973014 and the 973 Project (No. 2010CB833004).

\newcommand\ARNPS[3]{~Ann. Rev. Nucl. Part. Sci.{\bf ~#1}, #2~ (#3)}
\newcommand\AL[3]{~Astron. Lett.{\bf ~#1}, #2~ (#3)}
\newcommand\AP[3]{~Astropart. Phys.{\bf ~#1}, #2~ (#3)}
\newcommand\AJ[3]{~Astron. J.{\bf ~#1}, #2~(#3)}
\newcommand\APJ[3]{~Astrophys. J.{\bf ~#1}, #2~ (#3)}
\newcommand\APJL[3]{~Astrophys. J. Lett. {\bf ~#1}, L#2~(#3)}
\newcommand\APJS[3]{~Astrophys. J. Suppl. Ser.{\bf ~#1}, #2~(#3)}
\newcommand\JHEP[3]{~JHEP{\bf ~#1}, #2~(#3)}
\newcommand\JCAP[3]{~JCAP {\bf ~#1}, #2~ (#3)}
\newcommand\LRR[3]{~Living Rev. Relativity. {\bf ~#1}, #2~ (#3)}
\newcommand\MNRAS[3]{~Mon. Not. R. Astron. Soc.{\bf ~#1}, #2~(#3)}
\newcommand\MNRASL[3]{~Mon. Not. R. Astron. Soc.{\bf ~#1}, L#2~(#3)}
\newcommand\NPB[3]{~Nucl. Phys. B{\bf ~#1}, #2~(#3)}
\newcommand\PLB[3]{~Phys. Lett. B{\bf ~#1}, #2~(#3)}
\newcommand\PRL[3]{~Phys. Rev. Lett.{\bf ~#1}, #2~(#3)}
\newcommand\PR[3]{~Phys. Rep.{\bf ~#1}, #2~(#3)}
\newcommand\PRD[3]{~Phys. Rev. D{\bf ~#1}, #2~(#3)}
\newcommand\RMP[3]{~Rev. Mod. Phys.{\bf ~#1}, #2~(#3)}
\newcommand\SJNP[3]{~Sov. J. Nucl. Phys.{\bf ~#1}, #2~(#3)}
\newcommand\ZPC[3]{~Z. Phys. C{\bf ~#1}, #2~(#3)}
 \newcommand\IJGMP[3]{~Int. J. Geom. Meth. Mod. Phys.{\bf ~#1}, #2~(#3)}
  \newcommand\GRG[3]{~Gen. Rel. Grav.{\bf ~#1}, #2~(#3)}

\end{document}